\documentstyle[12pt]{article}
\newcommand{\bm}[1]{\mbox{\boldmath $#1$}}

\input{psfig}
\title{Resonant enhanced diffusion in time dependent flow}
\author{P.~Castiglione$^{1}$, A.~Crisanti$^{1}$, A.~Mazzino$^{2,3}$,\\ 
M.~Vergassola$^{2}$ and A. Vulpiani$^{1}$\\
\small $^{1}$ 
 Dipartimento di Fisica, and Istituto Nazionale di Fisica della Materia,\\
\small Universit\`a ``La Sapienza'', P.le A. Moro 2, 00185
Roma, Italy.\\
\small $^2$ CNRS, Observatoire de Nice, B.P. 4229,
06304 Nice Cedex 4, France.\\
\small $^{3}$ 
Dipartimento di Fisica, and Istituto Nazionale di Fisica della Materia,\\
\small Universit\`a di Genova, Via Dodecaneso 33, 16146
Genova, Italy.}

\begin{document}
\maketitle
\date{}

\begin{abstract}

Explicit examples of scalar enhanced diffusion due to resonances
between different transport mechanisms are presented. Their signature
is provided by the sharp and narrow peaks observed in the effective
diffusivity coefficients and, in the absence of molecular diffusion,
by anomalous transport. For the time-dependent flow considered here,
resonances arise between their oscillations in time and either molecular
diffusion or a mean flow. The effective diffusivities are calculated
using multiscale techniques.

\end{abstract}
PACS number(s): 47.27Qb\,; 47.27-i.\\

\section{Introduction}

    Passive scalar transport in a given velocity field is an issue of
    both theoretical and applicative relevance \cite{HKM,NCim}.  The
    combination of molecular and advective effects can lead to rather
    subtle behaviors, even in the case of simple laminar flow. A
    rather complete theory has been developed for the dynamics at long
    times and large scales (macrodynamics), using multiscale
    techniques \cite{BLP78}.  For incompressible velocity fields the
    macrodynamics in the presence of scale separation is governed by
    an effective equation which is always diffusive and transport is
    always enhanced \cite{Piretal}. A sufficient condition ensuring
    the presence of scale separation for three--dimensional
    incompressible flow is a vector potential finite variance, both
    for static and time--dependent flow \cite{AM,AV} (with similar
    conditions for other space dimensionalities). The small--scale
    velocity field properties emerge via the effective diffusivity
    second--order tensor.  Its calculation is reduced to the solution
    of one auxiliary partial differential equation that can be
    efficiently solved numerically \cite{Majetal,BCVV95} and
    variational principles have also been derived \cite{FP}.  The
    effective diffusivities for generic flow depend on all the
    mechanisms of transport, e.g. on the advecting velocity field, on
    the molecular noise, on the presence and the strength of possible
    mean flow \cite{Majetal,M97}.  Each process is characterized by
    its typical time--scales and, if the latter are properly tuned,
    their mutual interference can produce resonance effects. These
    lead to strong enhancement of transport, reflected in sharp and
    narrow peaks in the effective diffusivity. Our aim here is to
    present explicit examples of such resonance effects in simple
    time--dependent incompressible velocity fields.  In Section 2 the
    source of resonance is the presence of a large-scale velocity
    field. For random parallel flow (Sec.~2.1), explicit analytic
    expressions for the effective diffusivities are obtained and they
    indeed display strong peaks for specific strengths of the
    large-scale velocity.  In Section ~2.2 the robustness of resonance
    effects to variations of the scale separation ratio between the
    large- and the small-scale velocity field components is
    investigated.  Numerical simulations show that the effects persist
    even for moderate scale separations.  Section~3 is devoted to a
    time-periodic $2D$ flow widely studied to mimic the transport in
    Rayleigh-B\'enard system \cite{GS88a,GS88b}. The snapshots of the
    velocity field are made of closed lines. The syncronization
    between the circulation in the cells and their global oscillation
    provides however for a very efficient way of jumping from cell to
    cell. This mechanism, similar to stochastic resonance
    \cite{BSV81,BPSV83}, makes the curve of the effective diffusivity
    {\it vs} the frequency $\omega$ of oscillation of the flow very
    structured. In the limit where the molecular diffusion vanishes,
    anomalous superdiffusion takes place for narrow windows of values
    of $\omega$ around the peaks.  The anomaly is observed both in the
    second--order moment of particle dispersion and in the singular
    behavior of the effective diffusivity at high P\'eclet numbers.
    In Section ~4 we study a $2D$ system consisting of the
    superposition of steady coherent structures and a random
    incoherent field with a typical correlation time $\tau$.  The
    resonance occurs now when the sweeping time of the coherent
    structure is close to $\tau$.  Conclusions are reserved for the
    final Section~5.

\section{Parallel flow with a large scale velocity field}
The passive scalar field $\theta({\bm x},t)$ obeys
(see Ref.~\cite{C43}) the equation\,:
\begin{equation}
\partial_t\theta(\bm{x},t)+\left (\bm{v}\cdot \bm{\nabla}\right )
\theta(\bm{x},t)=
D_0\Delta\theta (\bm{x},t). 
\label{FP}
\end{equation}
The velocity $\bm{v}(\bm{x},t)$ is here assumed to be incompressible
and given by the sum of two terms, $\bm{u}$ and $\bm{U}$.  The first
is periodic both in space (in a cell of size $l$) and in time (the
technique can be extended with some modifications to handle the case
of random, homogeneus and stationary velocity fields).  The second is
the large--scale component of $\bm{v}$, varying on a typical scale
$L$.  It is assumed that $l/L =\epsilon \ll 1$, where $\epsilon$ is
the parameter controlling the scale separation.

We are interested in the dynamics of the field $\theta (\bm{x},t)$ on
large scales $\sim O(L)$, comparable to those of ${\bm U}$.  In the
spirit of multiscale methods \cite{BLP78}, slow variables ${\bm
X}=\epsilon {\bm x}$, $T={\epsilon}^{2} t$ and
$\tau_{\epsilon}=\epsilon t$ are introduced in addition to the fast
variables ${\bm x}$ and $t$.  The time variable $\tau_{\epsilon}$ is
needed to account for the advective effects generated by the presence
of ${\bm U}$.  The effective equation governing the dynamics at large
scales of the field $\theta$ is \cite{M97}\,:
\begin{equation}
\label{eqeff}
\partial_t\theta+{\bm U}\cdot\bm{\nabla}
\theta=
\nabla_{\alpha}D_{\alpha\beta} \nabla_{\beta}\theta ,
\end{equation} 
where the second--order eddy--diffusivity tensor is given by 
\begin{equation}
D_{\alpha \beta}(\bm{X},T)=\delta_{\alpha \beta}D_0-{1\over 2}\left[ 
\langle u_{\alpha}w_{\beta} \rangle + \langle u_{\beta}w_{\alpha} \rangle
\right],
\label{meglio}
\end{equation}
and the symbol $\langle\cdot\rangle$ denotes the average over the
periodicities of the small--scale velocity field ${\bm u}$. The
auxiliary field $\bm{w}(\bm{ x},t ;\bm{ X},T)$ has vanishing average
over the periodicities and satisfies the equation
\begin{equation}
\label{chi}
\partial_t\bm{ w}+\left[(\bm{ u}+ \bm{ U})\cdot\bm{ \partial}\right]
\bm{ w} - D_0\, \partial^2\bm{ w} = -\bm{u} \;\;\; .
\end{equation} 

\subsection{A solvable case}

A well-known situation where the effective diffusivities can be
calculated analytically is the one of parallel flow. Let us consider
in detail the case of three-dimensional random parallel flow in the
presence of a large scale advecting velocity field ${\bm U}({\bm
X},T)$\,:
\begin{equation}
{\bm u}({\bm x},t)= \left ( u(y,z,t), 0, 0 \right )\,;\qquad
{\bm U}({\bm X},T)= \left (0, U(X,Z,T), 0 \right ) .
\end{equation}
The field ${\bm u}({\bm x},t)$ is random, homogeneus and
stationary. The specific dependence on the spatial variables makes the
fields automatically incompressible.  The above decomposition is
standard in the framework of me\-sos\-ca\-le meteorology \cite{P84}.
Its physical meaning is that small-scale eddies remain stationary
while slow modifications of the large scale component occur. This is
quite a common feature in geophysics, e.g. in the planetary boundary
layer \cite{S89}, where the airflow in the thin atmospheric layer near
the ground is strongly driven by sink/source forcing terms arising
from the bottom boundary.

The first component of the equation (\ref{chi}), the only one for which the 
r.h.s. does not vanish, can be easily solved in
Fourier space and its solution reads:
\begin{equation}
\hat{w}_1 ({\bm k},\omega;{\bm X},T)=
\frac{-\hat{u}({\bm k},\omega)}{i\omega + k^2 D_0 + i
{\bm U}\cdot {\bm k}}= -\hat{u}({\bm k},\omega)
\hat{G}({\bm k},\omega ;{\bm X}, T) \;\;\; ,
\end{equation}
where the advection--diffusion 
propagator $\hat{G}({\bm k},\omega ; {\bm X}, T) $ is defined as:
\[ \hat{G}({\bm k},\omega ; {\bm X}, T) =
\frac{1}{i\omega + k^2 D_0 + i{\bm U}\cdot {\bm k}}= \int_{0}^{+\infty}
e^{-[i\omega + k^2 D_0 + i{\bm U}\cdot{\bm k}]\alpha} d\alpha \;\;\; .\]

It follows from (\ref{meglio}) that the only non--vanishing elements of
the eddy--diffusivity tensor are diagonal and\,:
\begin{equation}
D_{xx}({\bm X},T)=D_0 +
 \int_0^{\infty} S(t) \int \hat{E}({\bm k})
e^{-[k^2 D_0 + i\bm{U}\cdot {\bm k} ] t }\; dt\; d{\bm k} \,;\ \ 
D_{yy}=D_{zz}=D_0 \;\; ,
\label{utile}
\end{equation}
We have for simplicity assumed the separability 
$\hat{E} ({\bm k},\omega)= \hat{E}({\bm k}) \hat{S}(\omega) $, where
$\hat{E}({\bm k})$ and $\hat{S}(\omega) $ are the 
spectra of the spatial and the temporal part of the correlation 
function of ${\bm u}$.

The simple analytic expression (\ref{utile}) can be used to address
the question of interest in this paper, i.e., whether it is possible
to have a resonance between the oscillation frequencies induced by $U$
and the small--scale velocity field. We shall assume for simplicity
$U$ to be independent of both $\bm{X}$ and $\bm{T}$ and the following
temporal part of the velocity correlation function\,:
\begin{equation}
S(t)=e^{-|t|/\tau} \cos(\omega t)\;\;\; .
\label{st}
\end{equation}
Substituting (\ref{st}) into (\ref{utile}) and performing a simple
integration, the latter takes the form:
\begin{equation}
D =D_0+ \tau_0
\int_0^{\infty} \hat{E}(k) R(k) \; d k .
\label{interfer}
\end{equation}
Here, the function $R(k)$ is
\begin{equation}
R(k) = 
\frac{\left (\frac{1+\omega^2 \tau^2 }{1+k^2 D_0 \tau}\right )
\left [1+
\left( \frac{\omega\tau}{1+k^2 D_0 \tau}\right )^2 + 
\left (\frac{U \tau k}{1+k^2 D_0\tau }\right )^2\right ]}
{
\left[1+
\left(\frac{\omega\tau}{1+k^2 D_0 \tau}\right )^2 +
\left ( \frac{U k \tau}{1+D_0 k^2 \tau} \right )^2
\right ]^2 -
4 \left [\frac{\omega  U k \tau^2}{\left (1+D_0 k^2\tau\right)^2 }\right ]^2}
\label{erre}
\end{equation}
and $\tau_0$ is the correlation time of turbulence:
\begin{equation}
\tau_0=\int_0^\infty S(t)\; dt = \frac{\tau}{1+ \omega^2 \tau^2} \;\;\; .
\label{autoc}
\end{equation}
Note that $R(k)\rightarrow 1$ when both $U \rightarrow 0$ and
$D_0\rightarrow 0$.  In the latter case $D=u_0^2 \tau_0$, the well
known result valid for velocity fields $\delta$--correlated in time
(see, e.g., Ref.~\cite{BCVV95}).  It follows from (\ref{interfer})
that the combined effect of $D_0$ and $U$ is to modify the correlation
time of turbulence, producing an effective correlation time
$\tau_{eff}=\tau_0 R(k)$, which depends on the wavenumber $k$.  It is
precisely such a dependence on $k$ that makes it possible to have
resonances. Let us indeed consider (\ref{erre}) for a vanishing value
of $D_0$. It is easy to verify that, for $\omega\tau > 1/\sqrt{3}$,
$R(k)$ has a peak for
\begin{equation}
U^{res} = 
\frac{1}{k \tau}\sqrt{-\omega^2 \tau^2 -1 
+ 2\omega\tau \sqrt{\omega^2\tau^2 +1} } .
\end{equation}
In the limit $\omega\tau \gg 1$, the previous expression reduces to\,:
$U^{res} = \frac{\omega}{k}$. Note that $R(k)$ crosses the unit value
for $\omega\tau > 1/\sqrt{3}$. The regime $R(k)>1$ corresponds to
constructive interference between the turbulent motion and the
slow--varying one \cite{MV97}. Destructive interference takes place,
on the contrary, for $\omega\tau \leq 1/\sqrt{3}$ where 
$R(k)$ monotonically decreases from $1$ ($U=0$) to zero (large $U$).
The effective diffusivity for small-scale flows with a single wavenumber 
$\bm{k}_0$, attains at the maximum the value 
\begin{equation}
D^{res}=u_0^2 \tau_0 
\frac{(1+\omega^2 \tau^2)(1+2\omega^2 \tau^2)}{1+4\omega^2 \tau^2}.
\end{equation}
When $\omega\tau \gg 1$ (and the velocity $U$ is `synchronized' with
the oscillating velocity field ($U^{res}=\omega/k_0$)), the previous
expression reduces to $u_0^2 \tau_0 \frac{\omega^2 \tau^2}{2}$.
For $U \gg \omega/k_0$, the effective diffusivity $D$ goes to zero 
quadratically, as $D=u_0^2 \tau_0\left ( 
\frac{\omega}{k_0 U}\right )^2$.
%----------------------------------
%%%%\psdraft  %cosi' lascia uno spazio bianco dove risiede la figura.
\begin{figure}
\begin{center}
\mbox{\psfig{file=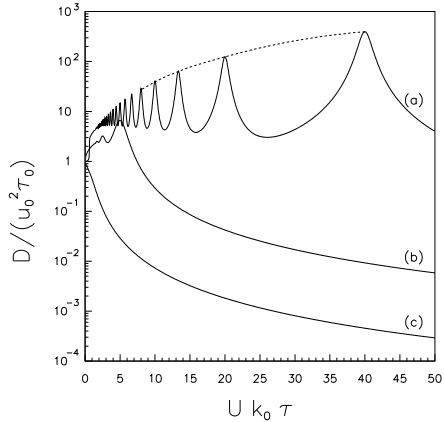,height=6cm,width=6cm}}
\end{center}
\caption{The eddy--diffusivity $D/(u_0^2\tau_0)$ {\em vs} the mean
flow strength $U k_0 \tau$. The curve is obtained by numerical
integration of (\protect\ref{interfer}) for $D_0=0$,
$\hat{E}(k)\propto k^{-5/3}$, temporal correlation function
$S(t)=e^{-|t|/\tau}\protect\cos(\omega t)$ and Strouhal number
$S\equiv u_0\tau_0 k_0\sim 1$. The three curves correspond to (a):
$\omega\tau=40$; (b): $\omega\tau=5$; (c):
$\omega\tau=1/\protect\sqrt{3}$.  The dashed line joins the peaks of
$D$ given by the expression (\protect\ref{picchi}) for $n=1,2,\cdots ,
5$.}
\label{para1}
\end{figure}
%%%%%%\psfull
%------------------------------------------------

The final result is that for a random parallel flow periodic in a box
of size $L$ ($k_0=2\pi/L$) and $\omega\tau \gg 1$, the peaks of the
effective diffusivity $D$ occur for
\begin{equation}
U_n^{res} = \frac{\omega}{k_0 n} \qquad n=1,2,\cdots ,  
\label{maxn}
\end{equation}
with values given, for large $U$, by :
\begin{equation}
D_n^{res} \sim \hat{E}(k_0 n) \tau_0 \frac{\omega^2 \tau^2}{2} \;\;\; .
\label{picchi}
\end{equation}
The behavior of $D$ as a function of $U$, obtained evaluating
numerically the integral (\ref{interfer}), is shown in
Fig.~\ref{para1} for a Kolmogorov spectrum $\hat{E}(k)\propto
k^{-5/3}$, and for three different values of $\omega\tau \geq
1/\sqrt{3}$.  Peaks of enhanced diffusion are evident for values of
$U$ given by (\ref{maxn}). Note that for $\omega\tau = 40$, the first
five peaks of $D$ (from right to left) are close to the values given
by
%------------------------------------------------
\begin{figure}
\begin{center}
\mbox{\psfig{file=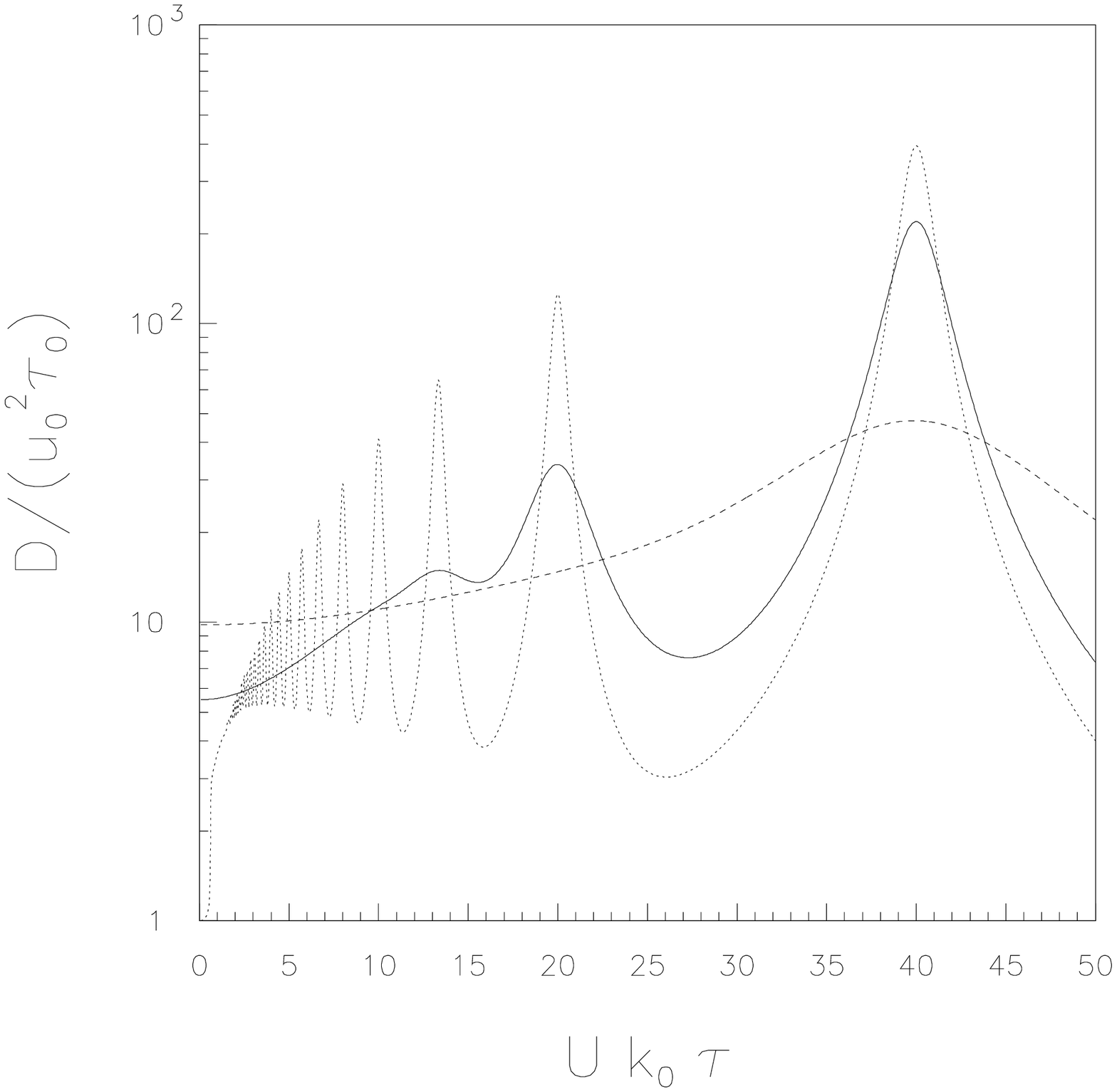,height=7cm,width=7cm}}
\end{center}
\caption{The same as in Fig.~\protect\ref{para1}, but 
the behavior of $D/(u_0^2\tau_0)$ is now plotted for 
three different values of $D_0$ and $\omega\tau=40$.
 Dotted line: $D_0=0$; full line: 
$D_0/(u_0^2\tau_0)=5\times 10^{-4}$; dashed line: 
$D_0/(u_0^2\tau_0)=5\times 10^{-3}$. Note that the larger is 
$D_0$, the broader are 
the peaks.}
\label{para2}
\end{figure}
%------------------------------------------------
the expression (\ref{picchi}).

The role of a non--vanishing molecular diffusivity $D_0$ 
is to reduce the height of the peaks. This point is shown 
in Fig.~\ref{para2}, where the behavior of $D/(u_0^2\tau_0)$  is 
plotted as a 
function of  $U k_0 \tau$, 
for $\omega\tau = 40$ and for three different values of $D_0$.
As we shall see in the following, these features hold also for  
more general flows than those considered in this section.  

It is finally worth noting that parallel flow permit to present yet
another simple example of resonance.  Considering indeed the
expression (\ref{interfer}) for vanishing values of $U$, it is easy to
verify that for
\begin{equation}
D_0=\frac{\omega\tau -1}{k^2\tau}\sim \frac{\omega}{k^2}
\end{equation}
peaks of $D$ occur. This corresponds to a synchronization between the
characteristic viscous frequency $k^2 D_0$ of the random molecular
noise and the frequency $\omega$ of the oscillatory velocity field.

\subsection{Parallel flow with moderate scale separations}

The aim here is to show that the resonant enhanced diffusion previously 
discussed also appears when only moderate scale separation factors are
present. As in (\ref{FP}), consider indeed the   
velocity field $ {\bm v}={\bm u}+{\bm U}$
with ${\bm u}$ parallel to the $x$ axis
\begin{equation} 
{\bm u} = \left( u_1\cos( k_1 y + \omega_1 t ), 0, 0 \right)
\label{incrocio1}
\end{equation}
and ${\bm U}$ parallel to the $y$ axis
\begin{equation}
{\bm U} = \left(0, u_2 \cos( k_2 x + \omega_2 t),0 \; 
\right) 
\label{incrocio2}
\end{equation}
and $k_1 > k_2$. The situation $k_1\gg k_2$ corresponds to the
scale-separated case previously considered. Here, we shall assume
$k_1=2.5\; k_2$ (and $u_1=u_2$) and we shall be interested in the
dynamics at scales much larger than $1/k_2$. The two components of the
velocity field ${\bm v}$ have moderate scale separation and affect
transport on equal footing.  The velocity ${\bm v}$ not being a
parallel flow, analytical expressions for the effective diffusivities
$D_{ij}$ are not available anymore. We had then to solve numerically
the standard auxiliary equation arising from multiscale methods (see,
e.g., Ref.~\cite{BCVV95}). The equation is solved by using a
pseudo-spectral method \cite{Gottl77} in the basic periodicity cell
with a grid mesh of $64 \times 64$ points.  De--aliasing has been
obtained by a proper circular truncation which ensures better isotropy
of numerical treatment.  Time marching has been performed using a
leap--frog scheme mixed with a predictor--corrector scheme (see
Ref.~\cite{March75}) at regular intervals.
 
\begin{figure}
\begin{center}
%\mbox{\psfig{file=fig11.eps}}
\mbox{\psfig{file=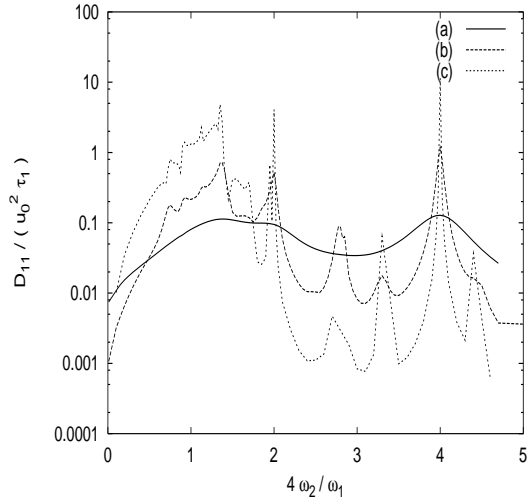,height=7cm,width=7cm}}
\end{center}
\caption{
The non-dimensional eddy--diffusivity $D_{11} /(u^2_0 \tau_1)$
{\em vs} the non-dimensional frequency $4 \omega_2 / \omega_1$, 
for different values of $D_0/(u^2_0 \tau_1)$. 
The cases ($a$), ($b$) and ($c$) correspond to 
$D_0/(u^2_0 \tau_1)=3 \times 10^{-3}$, 
$D_0/(u^2_0 \tau_1)=3 \times 10^{-4}$ and  
$D_0/(u^2_0 \tau_1)=3 \times 10^{-5}$, respectively.
}
\label{fig11}
\end{figure}

\begin{figure}
\begin{center}
%\mbox{\psfig{file=fig22.eps}}
\mbox{\psfig{file=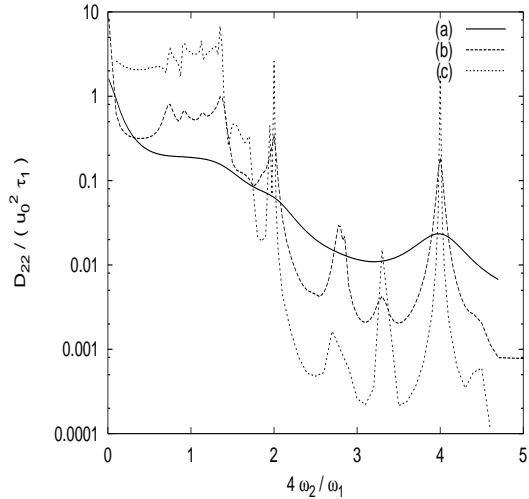,height=7cm,width=7cm}}
\end{center}
\caption{The non dimensional eddy--diffusivity $D_{22} /(u^2_0 \tau_1)$
{\em vs} the non dimensional frequency $4 \omega_2 / \omega_1$, 
for different values of $D_0/(u^2_0 \tau_1)$. 
The cases ($a$), ($b$) and ($c$) correspond to 
$D_0/(u^2_0 \tau_1)=3 \times 10^{-3}$, 
$D_0/(u^2_0 \tau_1)=3 \times 10^{-4}$ and  
$D_0/(u^2_0 \tau_1)=3 \times 10^{-5}$, respectively.
}
\label{fig22}
\end{figure}

To enlighten the resonant diffusion enhancement, $\omega_1$ is fixed
and the $D_{ij}$ behavior as a function of $\omega_2$ is investigated.
Results are reported in Figs.~\ref{fig11} and \ref{fig22}, where 
$D_{ii} /(u^2_0 \tau_1)$ {\it vs} $4 \omega_2 /\omega_1$
is presented for different values
of $D_0$. Here, $u^2_0=u^2_1+u^2_2$ and $\tau_1=2 \pi / \omega_1$ is
the oscillation period of the small-scale component.  Results clearly
exhibits resonance peaks for $\omega_2=\omega_1$, $\omega_2=\omega_1 /
2$, $\omega_2 \sim \omega_{sw}$ and $\omega_2 \sim 2 \omega_{sw}$,
where the frequency $\omega_{sw}$ is defined as:
$$\omega_{sw} = \frac{2 \pi}{T_{sw}}\qquad \mbox{with}\qquad T_{sw} =
\frac{L} {u_0 }$$ and $4 \; \omega_{sw} / \omega_1\; = \;
\sqrt{2}$. There are other peaks for $\omega < \omega_{sw}$,
corresponding to submultiples of $\omega_1$ and $\omega_{sw}$.

Note in Fig.~\ref{fig22} that $D_{22}$ increases for $\omega_2 \to 0$.
This is qualitatively understood as follows : for small $\omega_2 $,
the velocity field is essentially a stationary shear flow in the
$y$ direction superimposed to a rapidly changing small-scale
component.  The picture is rather close to the Taylor mechanism for
the longitudinal diffusion in the shear flow, which is well known to
lead to a $1/D_0$ behavior for $D_{22}$ \cite{new}.
     
%%%%\psdraft  %cosi' lascia uno spazio bianco dove risiede la figura.
%%%%%%\psfull

%%%%%%\psfull

%------------------------------------------------

%\section{Time periodic $2D$ flow}

%In this section we shall consider more complex and realistic flows.

\section{Diffusion in time-periodic Rayleigh-Benard convection}

Let us consider the two-dimensional velocity field defined by the
following stream function\,:
\begin{equation}
\psi (x,y,t) = \psi_0 \sin (x + B\sin\omega t ) \sin y \;\;\; .
\label{gollu}
\end{equation}
This flow ~\cite{GS88a,GS88b} is a simple model for transport in
time--periodic Rayleigh--B\'enard convection.  The stream function
(\ref{gollu}) describes single--mode, two--dimensional convection with
rigid boundary condition. The even oscillatory instability \cite{CB74}
is accounted for by the term $B\sin \omega t$, representing the
lateral oscillation of the roll.  In Ref.~\cite{GS88b}, a quantitative
comparison of the behavior in this flow with the experimental data has
shown that the basic mechanisms of convective transport are well
captured by the expression (\ref{gollu}).

For a fixed $\omega t$, the oscillation frequency of a Lagrangian
particle close to the center of the periodic cell is of the order of
$\psi_0$.  The periodicity of the cell is $L\equiv 2\pi$ and the
dimensionless parameter controlling the dynamics is $\epsilon =\omega
L^2 /\psi_0$.  The two limiting regimes $\epsilon \ll 1$ and $\epsilon
\gg 1$ have been investigated in Ref.~\cite{CNRY91} to obtain the
expressions of the diffusion coefficients in the limit of zero
molecular diffusivity. Here, we shall concentrate on the behavior for
$\epsilon\sim 1$. It is indeed precisely in this region that a
synchronization between the characteristic frequencies $O(\psi_0)$ and
$O(\omega)$ can take place.  In all the cases reported here, a
resolution $ 512 \times 512 $ has been used, which is found adequate
for molecular diffusivities $D_0/\psi_0\geq 5\times 10^{-4}$.  The
system evolution has been computed up to times $500\;L^2/\psi_0$,
with a time step $\Delta t = 10^{-2} L^2/\psi_0$.

For small enough (or vanishing) numerical diffusivities, the numerical
integration of the partial differential equation (\ref{chi}) is
clearly subject to instabilities. To evaluate the eddy--diffusivities,
it is then more convenient to integrate the equation $d_t{\bm x}={\bm
v}$ using a second-order Runge-Kutta scheme and then performing a
linear fit of $\langle [x(t) - x(0)]^2\rangle$ {\it vs } $t$, to obtain
the diffusion coefficient. The averages are made over different
realizations and are performed by uniformly distributing $10^6$
particles in the basic periodic cell.  The system evolution has been
computed up to times $10^4\;L^2/\psi_0$.  Numerical schemes have been
tested both for vanishing values of $\omega$ and for large P\'eclet
number. Asymptotic solutions for the effective diffusivity are indeed
available in these limits (see Eq. (27) in Ref.~\cite{RBDH87}).
Measured diffusion coefficients agree with the theoretical values within
errors of less than $1\%$.
%-----------------------------------------------------------------
\begin{figure}
\vfill \begin{minipage}{.495\linewidth}
\begin{center}
\mbox{\psfig{figure=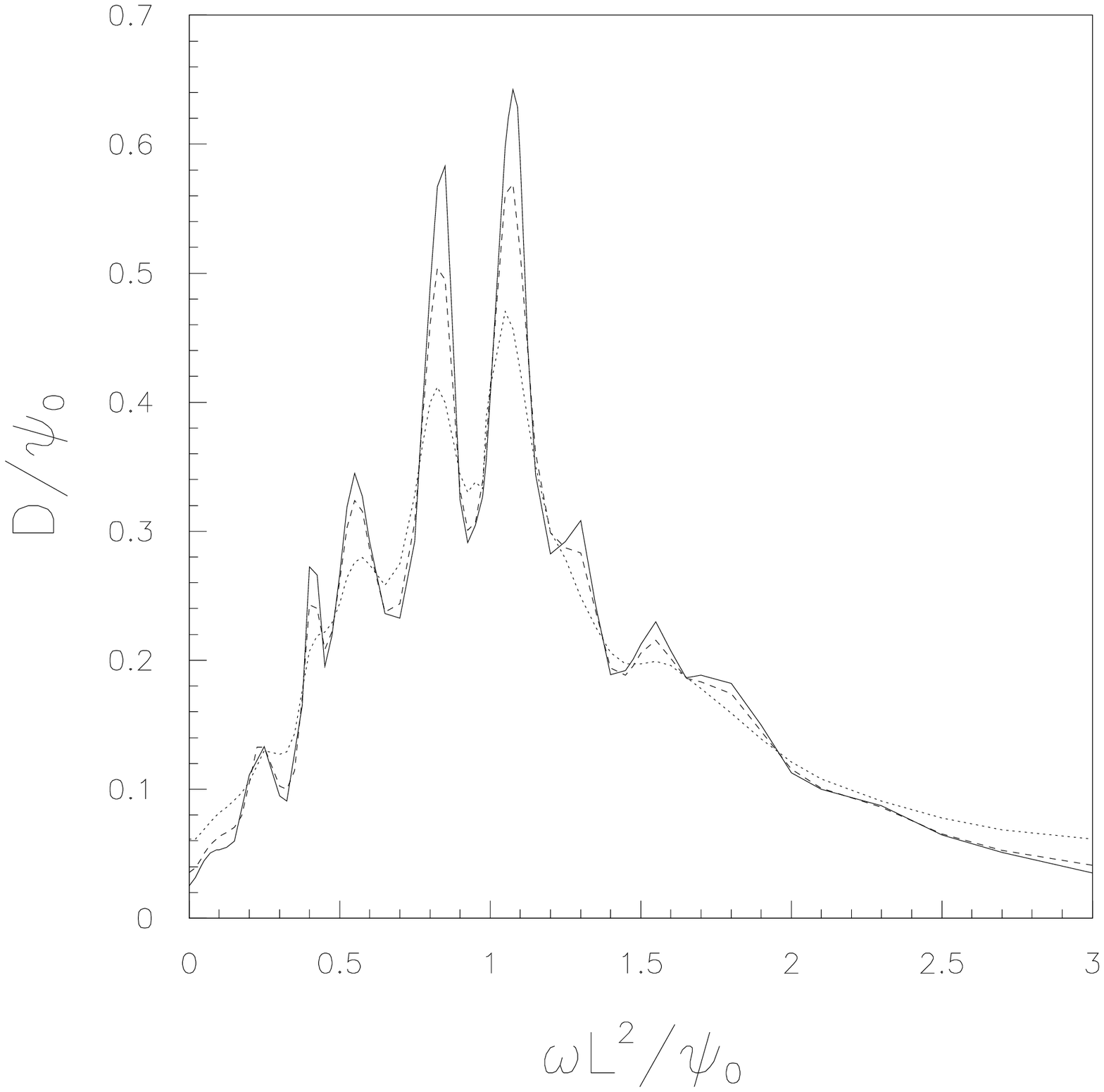,width=.9\linewidth}}
\end{center}
\end{minipage} \hfill
\begin{minipage}{.495\linewidth}
\begin{center}
\mbox{\psfig{figure=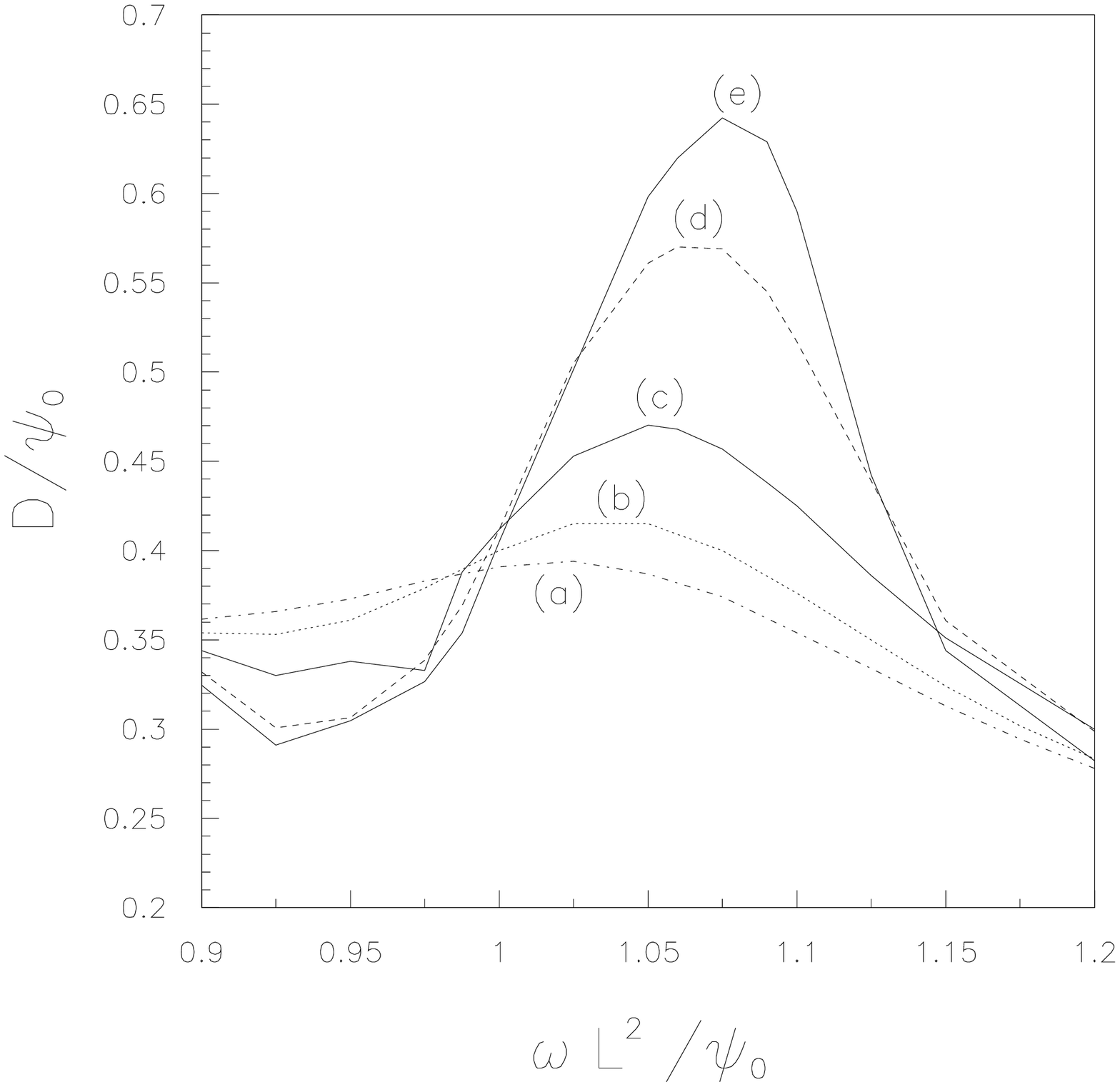,width=.9\linewidth}}
\end{center}
\end{minipage}
\caption{The turbulent diffusivity $D/\psi_0$ {\it vs} 
the frequency $\omega L^2 /\psi_0$ for different values of the 
molecular diffusivity $D_0/\psi_0$. 
On the left: $D_0/\psi_0=3\times 10^{-3} $ (dotted line);
$D_0/\psi_0=1\times 10^{-3} $(dashed line); $D_0/\psi_0=5\times 10^{-4} $ 
(full line).
On the right: 
$D_0/\psi_0=1\times 10^{-2} $ (a);
$D_0/\psi_0=6\times 10^{-3} $ (b); $D_0/\psi_0=3\times 10^{-3} $ (c); 
$D_0/\psi_0=1\times 10^{-3} $ (d); $D_0/\psi_0=5\times 10^{-4} $ (e). }
\label{spike}
\end{figure}
%--------------------------------------------------

The turbulent diffusivity (the $x$-component) {\it vs} the frequency
$\omega L^2/\psi_0$ is shown in Fig.~\ref{spike} for different values
of $D_0/\psi_0$.  A few comments are in order.  First, the turbulent
diffusivity shows maxima which seems similar to those observed for
stochastic resonance \cite{BSV81, BPSV83}. Here, the resonance is
between the lateral roll oscillation frequency and the characteristic
frequencies of the scalar field motion.  Second, the effect of $D_0$
on the shape of the peaks is twofold\,: similarly to the case of the
parallel flow, the smaller is $D_0$, the sharper are the peaks.
Furthermore, from the expanded view of Fig.~\ref{spike} (on the
right), it appears that variations of $D_0$ also cause a shifting of
maxima positions.
%------------------------------------------------
\begin{figure}
\begin{center}
\mbox{\psfig{file=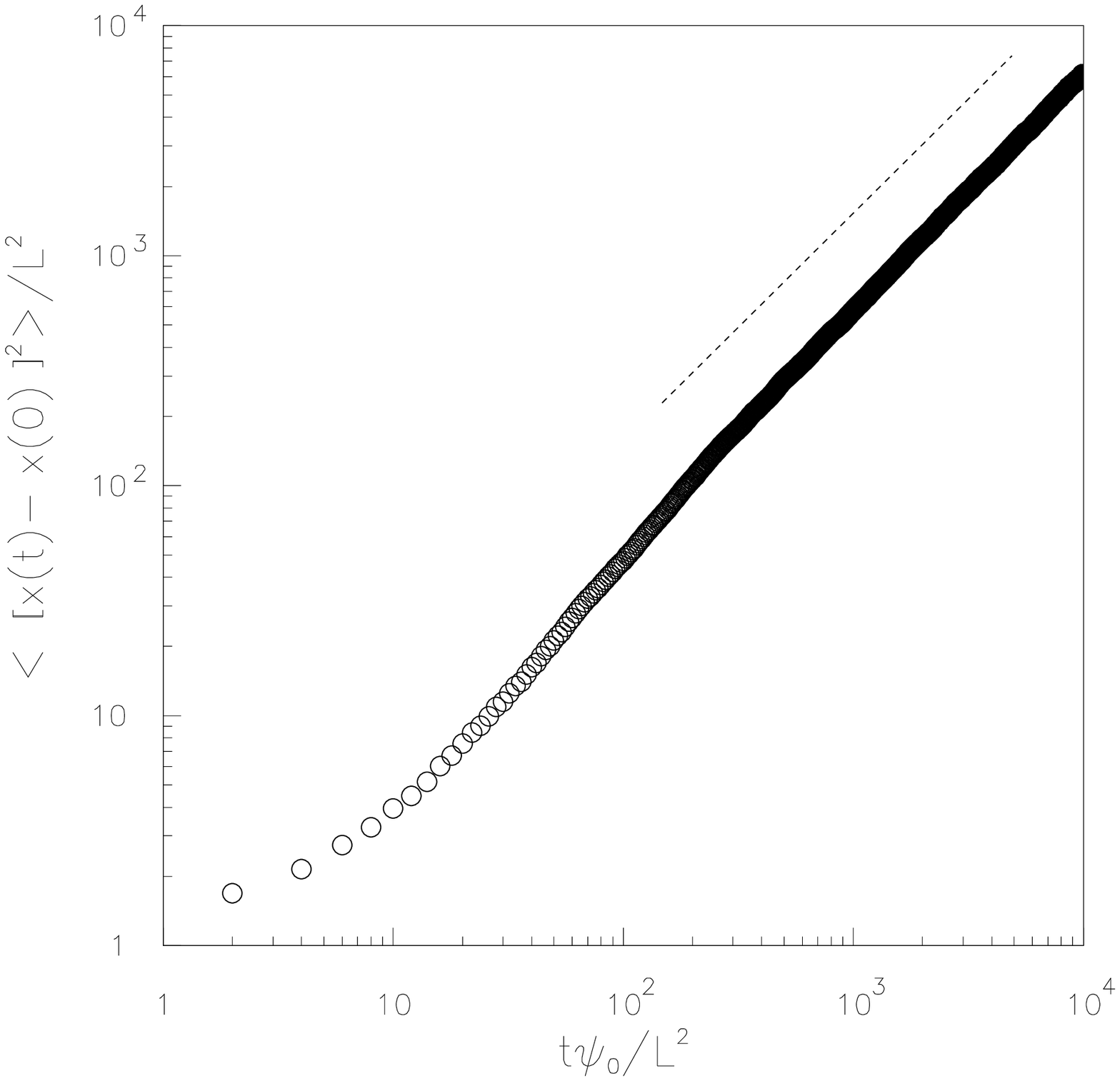,height=5.5cm,width=5.5cm}}
\end{center}
\vspace{-1.2cm}
\begin{center}
\mbox{\psfig{file=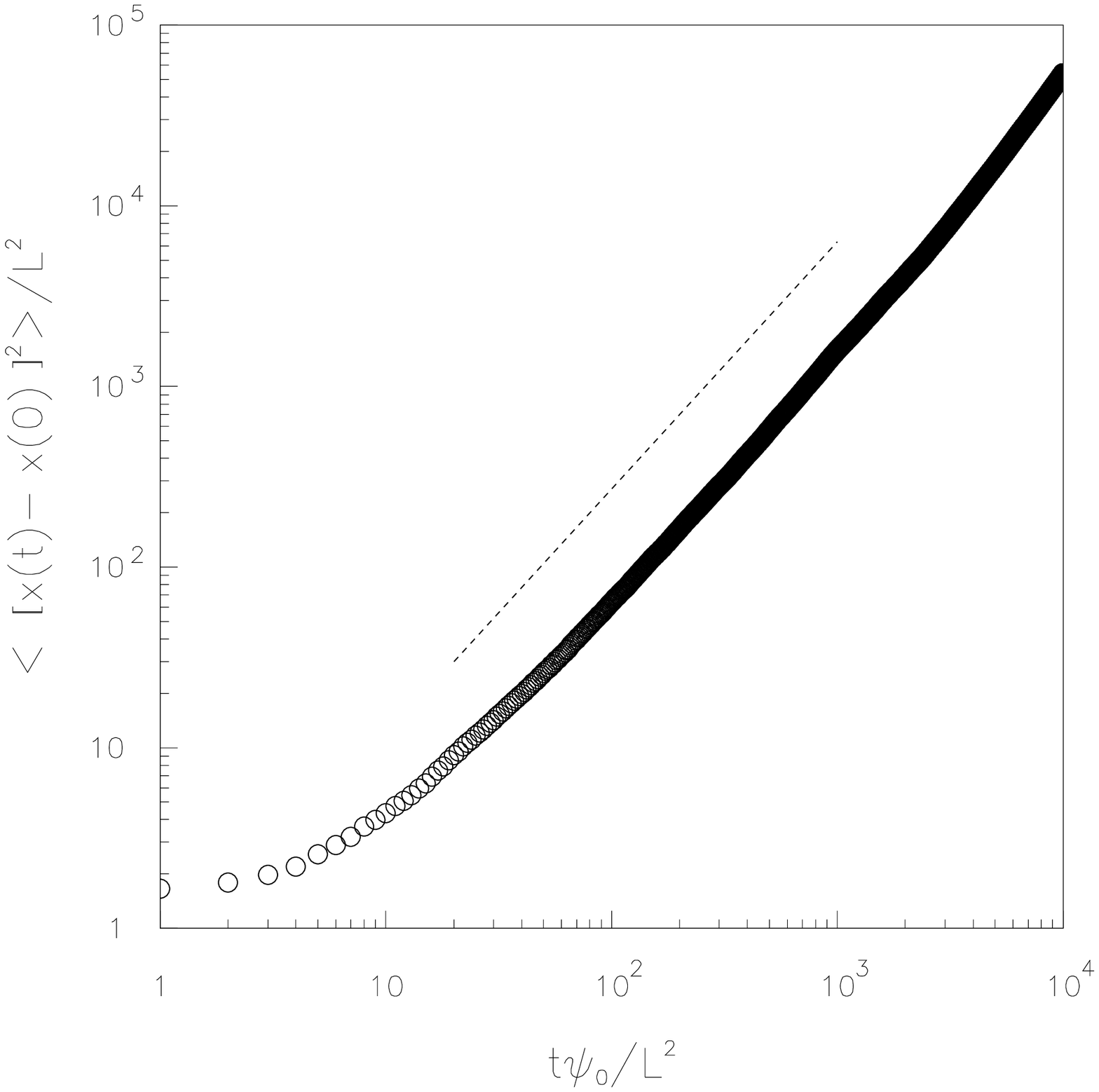,height=5.5cm,width=5.5cm}}
\end{center}
\vspace{-1.2cm}
\begin{center}
\mbox{\psfig{file=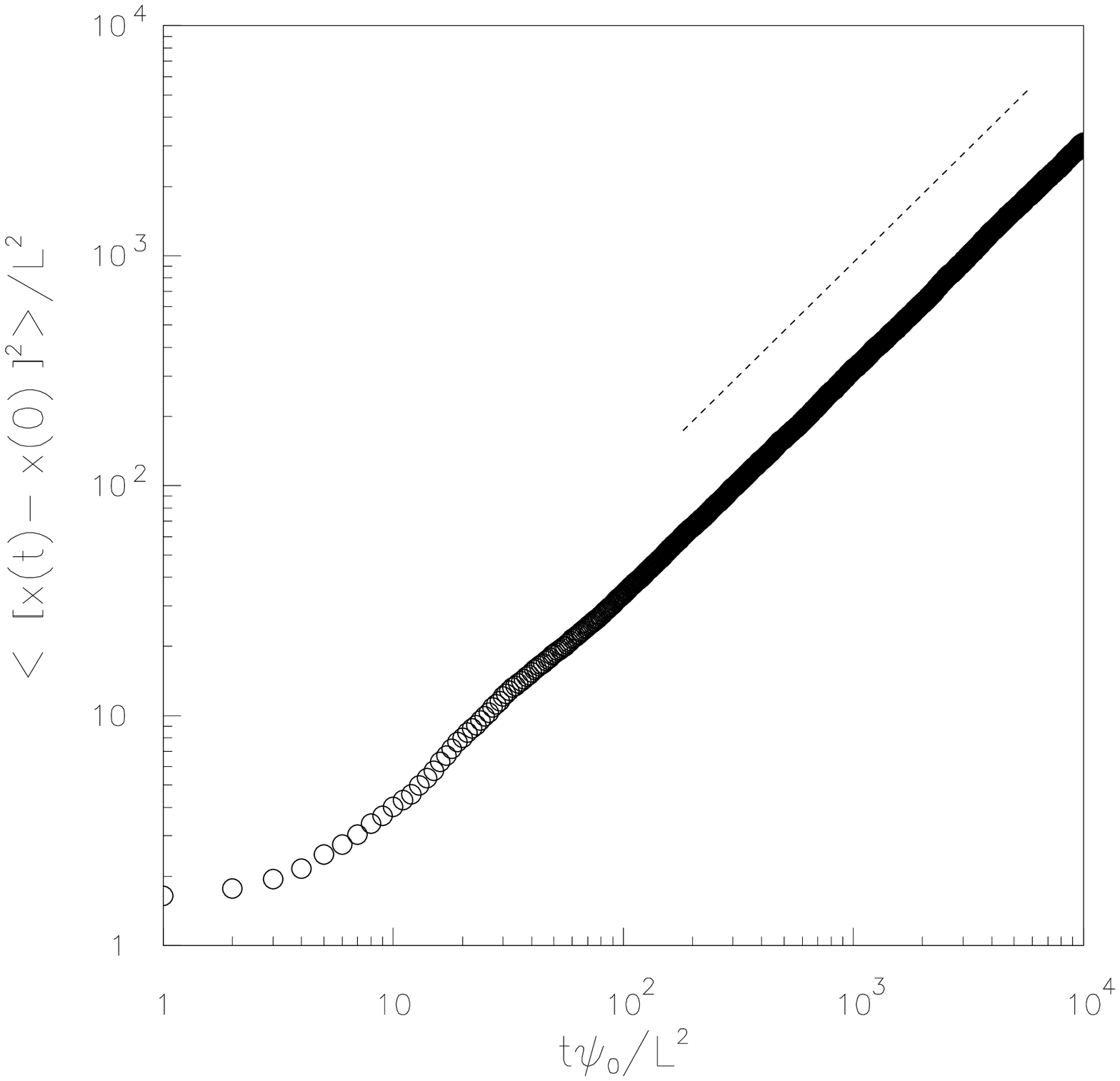,height=5.5cm,width=5.5cm}}
\end{center}
\caption{The mean--squared displacement {\em vs} the time for the flow
(\protect\ref{gollu}) with $D_0=0$ (in a log-log scale). From above to
below: $\omega L^2/\psi_0 = 1.075$, $\omega L^2/\psi_0 = 1.1$ and
$\omega L^2/\psi_0 = 1.15$.  Dashed lines have slopes $1$, $1.4$ and
$1$, respectively.}
\label{anom}
\end{figure}
%------------------------------------------------

For $D_0\to 0$, the peak shown in Fig.~\ref{spike} moves toward
$\omega L^2/\psi_0\sim 1.1$ and the transport becomes superdiffusive
at $D_0=0$, i.e. the mean--squared displacement is superlinear:
\begin{equation}
\langle   [x(t) - x(0) ]^2\rangle \propto t^{\alpha} \qquad \mbox{with}\qquad 
\alpha > 1\;\;\; . 
\label{anomalo}
\end{equation}
The range of $\omega L^2/\psi_0$ values where the process is
superdiffusive turns out to be very narrow: for $\omega L^2/\psi_0 =
1.1$ the mean--squared displacement behavior is given by
(\ref{anomalo}) with exponent $\alpha\sim 1.4$; both for $\omega
L^2/\psi_0 = 1.075$ and $\omega L^2/\psi_0 = 1.15$ the diffusion
process is standard (i.e., $\alpha = 1$).  This point is illustrated
in Fig.~\ref{anom}, where the mean--squared displacement as a function
of the time is shown in a log--log plot for : $\omega L^2/\psi_0 =
1.075$, $\omega L^2/\psi_0 = 1.1$ and $\omega L^2/\psi_0 = 1.15$.
%------------------------------------------------
\begin{figure}
\begin{center}
\mbox{\psfig{file=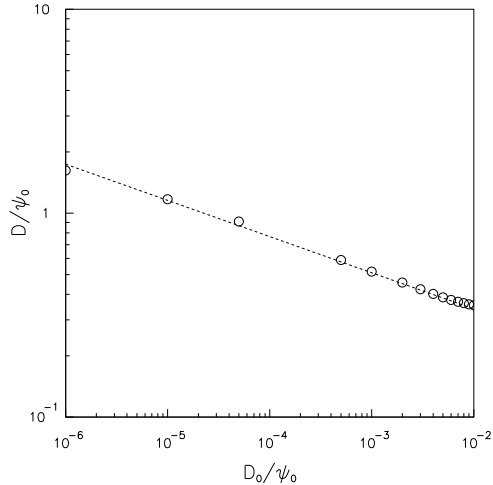,height=7cm,width=7cm}}
\end{center}
\caption{The diffusion coefficient $D$ as a function of $D_0$ at
$\omega L^2/\psi_0=1.1$ (in a log-log scale).  The slope of the dashed
line is $-\beta\sim 0.18$.}
\label{D_VS_D0_OM_1_1}
\end{figure}
%------------------------------------------------

    A more careful test of the anomalous diffusion at $D_0=0$ consists
    in computing the diffusion coefficient $D$ as function of
    $D_0$. As suggested in Ref.~\cite{BCVV95}, in the presence of
    genuine anomalous diffusion, the effective diffusivity must
    diverge and it is expected that $D \sim D_0^{-\beta}$ with $\beta
    > 0$. The curve $D$ versus $D_0$ at $\omega L^2/\psi_0=1.1$ is
    shown in Fig. \ref{D_VS_D0_OM_1_1}. The data are well fitted by a
    straight line with slope $\beta\sim 0.18$, confirming the presence
    of anomalous diffusion at $D_0=0$.

\section{Enhanced diffusion in the presence of coherent structures}

\begin{figure}
\begin{center}
\mbox{\psfig{file=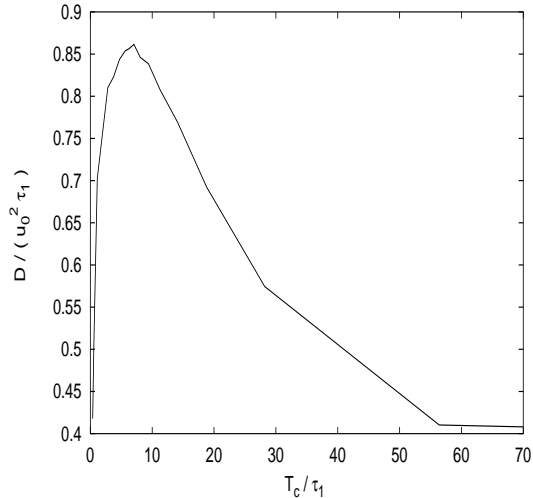,height=7cm,width=7cm}}
%\mbox{\psfig{file=vor.eps}}
\end{center}
\caption{The non dimensional eddy--diffusivity $D / (u_0^2 \tau_1 )$
{\em vs} the non dimensional Lagrangian time $T_c / \tau_1$ for $D_0
/(u_0^2 \tau_1) = 3.9 \times 10^{-4}$ and $k^I_M/k^I_m=4.5$.}
\label{vort}
\end{figure}

We shall consider now the case where the two-dimensional large-scale
component ${\bm U}$ consists of a coherent structure. The small-scale
part ${\bm u}$ is random.  Specifically, ${\bm u} = \left( u_1 , u_2
\right)$, where the $u_i$'s are homogeneous, isotropic and stationary
Gaussian random processes. Their Fourier transforms $\hat{u}_i$ have
zero mean and correlation functions :
\begin{equation} 
\langle \hat{u}_i({\bm k},t) \hat{u}_j ({\bm k}',t') \rangle =
(2 \pi)^2  \delta ({\bm k}+{\bm k}') 
\hat{R}_{ij} ({\bm k}) T({\bm k},t-t')
\label{corre}
\end{equation}   
where 
$$\hat{R}_{ij} ({\bm k}) =  (\delta_{ij}  k^2 - k_i  k_j )  f(k) 
\;\;\;\; \; \mbox{\rm and} \;\;\;\; 
T({\bm k},t-t')= e^{ -\frac{\mid t-t' \mid }{\tau(k)}}.
$$
The $f(k)$ function is related to the field energy spectrum $E(k)$ 
as follows:  
$$ E(k)=\frac{1}{2 \pi} k^3 f(k).$$ The following expressions for
$E(k)$ and $\tau(k)$ have been considered\,:
\begin{equation}
E(k) = \left\{
\begin{array}{c}
A \; k^{3}      \;\;\;\; k \in [ 0 , k^I_m ] \\   
B \; k^{-\frac{5}{3}}      \;\;\;\; k \in [ k^I_m , k^I_M ] \\
C \; e^{-k}      \;\;\;\; k > k^I_M \;\;\; .
\end{array}
\right.
\label{spect}
\end{equation}
and 
\begin{equation}
\tau(k) = \left\{
\begin{array}{c}
D \; k^{-3}      \;\;\;\; k \in [ 0 , k^I_m ] \\   
F \; k^{-\frac{2}{3}}      \;\;\;\; k \in [ k^I_m , k^I_M ] \\
G \; k^{-\frac{3}{2}} \; e^{\frac{k}{2}}      \;\;\;\; k > k^I_M 
\end{array}
\right.
\end{equation}
where the constants $A$, $C$ and $D$, $G$ are chosen to ensure the
continuity of $E(k)$ and $\tau(k)$ in $k^I_m$ and $ k^I_M$,
respectively. The expression for $\tau(k)$ is meant to mimic $3D$
turbulence, neglecting intermittency and correlations between
phases. The $\hat{u}_i({\bm k},t)$ can be easily generated by a linear
Langevin equation. The coherent large-scale component has the cellular
structure
\begin{equation}
{\bm U}= U \left( \cos y , \cos x \right)
\label{vortex}
\end{equation}
characterized by the typical Lagrangian time $T_c=\pi / (\sqrt{2} U)$.
>From the isotropy of the field it follows $D_{11}=D_{22}\equiv D$.  In
Fig.~\ref{vort} the diffusion coefficient $D/(u^2_0 \tau_1)$ is shown
as a function of $T_c/ \tau_1$, where $\tau_1=\tau(k^I_m)$ is a
characteristic time scale of the random field and $ u_0^2 = \int dk
\; E(k).$

The above results have been obtained performing direct numerical
simulations of the auxiliary equation (\ref{chi}) (now ${\bm U}=0$ and
${\bm u}$ has the spectrum defined by (\ref{spect})) by using a
pseudo-spectral method over the periodic box $2 \pi \times 2 \pi$ with
a resolution $ 64 \times 64 $.

It is well evident the presence of a peak for $T_c/\tau_1 \sim 7 $ and
$D_0 /(u_0^2 \tau_1) = 3.9 \times 10^{-4}$, whose value is $D / (u_0^2
\tau_1 )\sim 0.86$.  The fact that the peak does not appear for
$T_c/\tau_1$ of order unity is most likely due just to the freedom in
the definition of the characteristic times $T_c$ and $\tau_1$.  On the
contrary, it is important to note that the value of the peak is much
larger than the one occurring when either the random field or the
coherent structure are absent. The value $D / (u_0^2 \tau_1 )\sim
0.86$ has in fact to be compared with $0.07$ and $0.32$, respectively.

Unlike the resonance in the Reyleigh-B\'enard convection, the
syncronization mechanism produces here a broad peak.  This is not
surprising as in the present case the random field has a wide spectrum
of caracteristic times.

%%%%\psdraft  %cosi' lascia uno spazio bianco dove risiede la figura.

%%%%%%\psfull

\section{Conclusions}

Multiscale techniques have been used to present a series of explicit
examples of strong diffusion enhancement due to resonance effects
between different transport mechanisms.  The effects appear already
from the analytical expression of eddy-diffusivity for random flows
with a large scale mean flow component, are robust to a reduction of
the scale separation and they are also observed in convective
transport and when coherent structures are present on the large
scales.  For vanishing molecular diffusivities and for specific values
of the control parameters, the enhanced diffusion might turn into
anomalous superdiffusion.  Let us finally stress in which respect the
resonant enhanced diffusion effects discussed in this paper differ
from other known mechanisms to produce strong transport. For Taylor
longitudinal diffusion in shear flow, open streamlines producing large
effective diffusivities are for example present in the snapshots of
the velocity field itself. For the convective flow in Section~3,
snapshots of the velocity field are made at every time of closed
cells.  Open streamlines, and more generally enhanced transport, have
a dynamical origin and require a temporal syncronization between
different transport mechanisms.

\newpage
\section*{Acknowledgments}
We thank the ``Meteo--Hydrological Center of Liguria Region''
and the ``Department of Structural and Geotechnical Engineering 
of the University of Genova'' where part of the numerical
analysis was done.
AM is grateful to C.F. Ratto for very useful discussions and suggestions.
AM was supported by the ``Henry Poincar\'e'' fellowship 
(Centre National de la Recherche Scientifique and Conseil G\'en\'eral des
Alpes Maritimes) and by a fellowship ``di Prosecuzione'' (Universit\`a degli
Studi di Genova).  PC is grateful to the European Science Foundation for 
the TAO exchange grant. MV was supported by 
the GdR ``M\'ecanique des Fluides G\'eophysiques et Astrophysiques''.
PC, AC and AV were supported by INFM (PRA-TURBO).

%%%%% FINEEEEE

\end{document}